\documentclass[11pt,a4paper]{article}
\usepackage{amssymb}
\usepackage{times,cite}
\newcommand{\be}{\begin{eqnarray}}
\newcommand{\ee}{\end{eqnarray}}
\newcommand{\bez}{\begin{eqnarray*}}
\newcommand{\eez}{\end{eqnarray*}}
\newcommand{\pa}{\partial}
\newcommand{\la}{\lambda}
\newcommand{\ci}{\circ}

\newcommand{\A}{\mathbb{A}}

\title{\bf With a Cole-Hopf transformation to solutions of the 
     noncommutative KP hierarchy in terms of Wronski 
     matrices\thanks{\copyright
       2007 by A. Dimakis and F. M\"uller-Hoissen} }

\author{Aristophanes Dimakis \\
 Department of Financial and Management Engineering, \\
 University of the Aegean, 31 Fostini Str., GR-82100 Chios, Greece \\
 dimakis@aegean.gr
          \and
 Folkert M\"uller-Hoissen \\ 
 Max-Planck-Institute for Dynamics and Self-Organization \\
 Bunsenstrasse 10, D-37073 G\"ottingen, Germany \\
 folkert.mueller-hoissen@ds.mpg.de }

\date{}

\begin{document}

\maketitle

\begin{abstract}
In case of the KP hierarchy where the dependent variable takes 
values in an (arbitrary) associative algebra $\mathcal{R}$, it is known 
that there are solutions which can be expressed in terms of quasideterminants 
of a Wronski matrix which solves the linear heat hierarchy. 
We obtain these solutions without the help of quasideterminants in 
a simple way via solutions of matrix KP hierarchies (over $\mathcal{R}$) 
and by use of a Cole-Hopf transformation. For this class of exact 
solutions we work out a correspondence with `weakly nonassociative' algebras.  
\end{abstract}

The generalization of the KP equation to the case where the dependent variable 
takes values in a matrix algebra has already been considered long time ago in \cite{Zakh+Kuzn86,Atho+Ford87}, for example. The interest in this equation, 
and more generally in `soliton equations' where the dependent variable takes 
values in any associative algebra (see also \cite{Olve+Soko98assoc}, and in 
particular \cite{Dorf+Foka92,Kupe00} for the KP case), 
is partly due to the fact that there is an elegant way to 
generate from simple solutions of such a matrix or operator equation complicated 
solutions of the corresponding scalar equation   \cite{March88,Aden+Carl96,Carl+Schi99,Blohm00,Carl+Schi00,Schie02,Schie05,Han+Li01,Huan03,DMH06nahier,DMH06Burgers}. 
Moreover, certain developments in string theory motivated the study 
of soliton equations like the KP equation with the ordinary product of functions 
replaced by a noncommutative (Groenewold-Moyal) star product (see 
\cite{DMH04hier,DMH04ncKP,DMH05KPalgebra,Hama06} and references cited 
therein).\footnote{In several publications on `Moyal-deformed' soliton equations 
the Moyal-product can be replaced almost completely by \emph{any} associative 
(noncommutative) product, since the specific properties of the Moyal-product 
are not actually used. The algebraic properties of such equations are then simply 
those of (previously studied) matrix versions of these equations. Exceptions 
are in particular \cite{DMH04hier,DMH04ncKP,DMH05KPalgebra} where enlarged 
hierarchies are considered which appear specifically in the Moyal-deformed case. 
Multi-soliton solutions of the (enlarged) potential KP hierarchy with Moyal-deformed 
product were obtained in \cite{DMH05KPalgebra} using a method which, in the 
commutative case, corresponds to the well-known `trace method' \cite{Okhu+Wada83} 
(see also \cite{Kupe00}, appendix A6, and \cite{Pani01}). }
\vskip.1cm

Surprisingly, many integrability features of the scalar KP equation 
and its hierarchy generalize in some way to the `noncommutative' version. 
In \cite{EGR97} (see also \cite{Hama06qd,Gils+Nimm07}) expressions for solutions
of the `noncommutative' potential KP (pKP) hierarchy were found in terms of 
quasideterminants \cite{Gelf+Reta91,Gelf+Reta92,GKLLRT95,Krob+Lecl95,EGR98,GGRW05}, thus 
achieving a close analogy with classical results for the `commutative' pKP hierarchy 
(see \cite{Dick03,Hiro04}, for example).\footnote{One 
link between integrable systems and quasideterminants is given by the fact that 
`noncommutative' Darboux transformations \cite{Matv+Sall91} can be compactly expressed 
in the form of a quasideterminant \cite{Gonc+Vese98,GGRW05,Nimm06,Gils+Nimm07}.} 
In this letter we recover these solutions in an elementary way without the use 
of quasideterminants, via the construction of solutions of matrix pKP hierarchies, 
where now the matrices have entries in the respective (noncommutative) associative 
algebra. Moreover, our analysis sheds light on the underlying structure from 
different perspectives. Section~1 identifies a Cole-Hopf transformation as a 
basic ingredient. Solutions of the `noncommutative' pKP hierarchy obtained in 
this way determine solutions of a certain system of ordinary differential equations. 
In section~2 we show that conversely solutions of this system determine solutions 
of the pKP hierarchy. This is achieved by use of results from a very general 
approach towards solutions of KP hierarchies, developed in \cite{DMH06nahier} 
(see also \cite{DMH07Ricc}). 
Section~3 explains why Wronski matrices enter the stage under 
certain familiar additional conditions (``rank 1 condition" and shift operator). 
Here we make closer contact with the recent work in \cite{Gils+Nimm07}. 
Section~4 shows that the associated system of ordinary differential equations 
can then be cast into the form of matrix Riccati equations and makes contact with 
the Sato theory \cite{Sato+Sato82}. Finally, section~5 contains some further remarks 
and an appendix draws some consequences for the `noncommutative' \emph{discrete} 
KP hierarchy.
\vskip.2cm

\noindent
\textbf{1. Cole-Hopf transformation for noncommutative pKP hierarchies and 
related systems of ordinary differential equations.} \\
We recall a result from \cite{DMH06Burgers} (see theorem~4.1). Although not 
explicitly stated there, it holds for elements of an arbitrary associative algebra 
$\mathcal{A}$ with identity element $I$, over a field $\mathbb{K}$ of characteristic 
zero. It is assumed that the elements depend smoothly or as formal power series 
on independent variables $\mathbf{t} = (t_1,t_2, \ldots)$. 
Let $\pa : \mathcal{A} \rightarrow \mathcal{A}$ 
be a $\mathbb{K}$-linear derivation which commutes with the partial derivatives
$\pa_{t_n}$.\footnote{The reader would not run into problems setting 
$\pa = \pa_{t_1}$ in the following, because of (\ref{heat_hier}) with $n=1$.}
\vskip.2cm

\textbf{Proposition~1.} 
\emph{If $X,Y \in \mathcal{A}$ solve the linear heat hierarchy 
\be
      X_{t_n} = \pa^n(X) \, , \qquad n=1,2,3,\ldots \, ,  \label{heat_hier}
\ee 
(where $X_{t_n} = \pa_{t_n}(X)$) and if  
\be
       \pa(X) = R X + Q Y    \label{genCH_cond}
\ee 
with constant\footnote{An element $Q \in \mathcal{A}$ is called \emph{constant} 
if it does not depend on the variables $t_n$ and satisfies $\pa(Q)=0$.} 
elements $Q,R \in \mathcal{A}$, then
\be
     \phi = Y X^{-1}    \label{genCH}
\ee
solves the pKP hierarchy in the algebra $\mathcal{A}$ with product 
$A \cdot B = AQB$.
}
\hfill $\square$
\vskip.2cm

\textbf{Remark.} A functional representation of the pKP hierarchy in $(\mathcal{A}, \cdot)$
is given by $\Omega(\mu) - \Omega(\mu)_{-[\la]} = \Omega(\la) - \Omega(\la)_{-[\mu]}$ 
with 
\be
   \Omega(\la) = \la^{-1}(\phi - \phi_{-[\la]}) 
      - (\phi - \phi_{-[\la]}) \cdot \phi - \phi_{t_1} 
\ee 
and the Miwa shift $\phi_{-[\la]}(\mathbf{t}) = \phi(\mathbf{t}-[\la])$ where  
$[\la] = (\la,\la^2/2,\la^3/3,\ldots)$ (see \cite{DMH06Burgers} and also the 
references cited therein). Expansion in powers of the indeterminates 
$\la,\mu$ generates the pKP hierarchy equations. The pKP hierarchy system is 
equivalent to 
\be
   \Omega(\la) = \vartheta - \vartheta_{-[\la]}   \label{pKP2}
\ee 
with some $\vartheta \in \mathcal{A}$. Under the assumptions of proposition~1, 
we have $\vartheta = \phi R$ (see the proof of theorem~4.1 in \cite{DMH06Burgers}). 
In this restricted case, the hierarchy equations are thus determined by 
\be 
    (\phi - \phi_{-[\la]})(\la^{-1} - Q \phi - R) - \phi_{t_1} = 0 \; .
\ee
\hfill $\square$
\vskip.2cm

The equation (\ref{genCH}) becomes a \emph{Cole-Hopf} transformation if 
\be 
         Y = \pa(X) \, ,   \label{CH}
\ee
in which case the condition (\ref{genCH_cond}) takes the form
\be
    (I - Q) \pa(X) = R X \; .     \label{CH_cond}
\ee

\textbf{Proposition~2.} 
\emph{If $X \in \mathcal{A}$ solves the linear heat hierarchy (\ref{heat_hier}) 
and (\ref{CH_cond}) with constant $Q,R \in \mathcal{A}$, then 
\be
   \mathcal{W}(i,j) = \pa^{i+1}(X) X^{-1} R^j \qquad i,j=0,1,\ldots
      \label{Wij}
\ee
satisfy
\be
    \mathcal{W}(i,j)_{t_n} = \mathcal{W}(i+n,j) - \mathcal{W}(i,j+n) 
  - \sum_{k=0}^{n-1} \mathcal{W}(i,k) \, Q \, \mathcal{W}(n-k-1,j) \; .
    \label{Weqs_R}
\ee
}
\textbf{Proof:} By induction (\ref{CH_cond}) leads to
\bez
    \pa^n(X) = R^n X + \sum_{k=0}^{n-1} R^k Q \pa^{n-k}(X) 
    \qquad  n=1,2, \ldots \; .
\eez
With its help and by use of (\ref{heat_hier}) one easily verifies 
(\ref{Weqs_R}). 
\hfill $\square$
\vskip.2cm

Note that $\mathcal{W}(0,0) = \phi$. Associated with this solution 
of the pKP hierarchy in the algebra $\mathcal{A}$ with product $A \cdot B = AQB$, 
we thus have, via (\ref{Wij}), a solution $\{ \mathcal{W}(i,j) \}$ of the 
system (\ref{Weqs_R}) of ordinary differential equations. 
In the following section we prove the converse: whenever we have a 
solution $\{ \mathcal{W}(i,j) \}$ of the system (\ref{Weqs_R}), then 
$\mathcal{W}(0,0)$ solves the pKP hierarchy (in the algebra $\mathcal{A}$ 
with product $A \cdot B = AQB$). 
\vskip.2cm

\noindent
\textbf{2. Weakly nonassociative algebras related to pKP solutions.} \\
Before recalling a central result from \cite{DMH06nahier} (see also \cite{DMH07Ricc}), 
we need some definitions. An algebra $(\A,\circ)$ (over a commutative 
ring) is called \emph{weakly nonassociative (WNA)} if it is not associative, 
but the associator\footnote{The associator is defined as 
$(a,b,c) = (a \circ b) \circ c - a \circ (b \circ c)$.} 
$(a,b \circ c,d)$ vanishes for all $a,b,c,d \in \A$. The middle nucleus 
$\A' = \{ b \in \A  \, | \, (a,b,c) = 0 \; \forall a,c \in \A \}$, which is an 
\emph{associative} subalgebra, is then also an ideal in $\A$. 
With respect to an element $\nu \in \A \setminus \A'$ 
we define the products $a \circ_1 b = a \circ b$ and 
\be
        a \circ_{n+1} b = a \circ (\nu \circ_n b) - (a \circ \nu) \circ_n b  
        \qquad \quad  n=1,2, \ldots \; .    \label{circ_n}
\ee
As a consequence of the WNA condition, these products only depend on the equivalence 
class $[\nu]$ of $\nu$ in $\A/\A'$.
\vskip.2cm

\textbf{Theorem} \cite{DMH06nahier}.
\emph{
Let $\A$ be any WNA algebra, the elements of which depend smoothly 
on independent variables $t_1,t_2, \ldots$, and let $\nu \in \A \setminus \A'$ 
be constant. Then the flows of the system of ordinary differential equations
\be
  \phi_{t_n} = - \nu \circ_n \nu + \nu \circ_n \phi + \phi \circ_n \nu 
      - \phi \circ_n \phi  \qquad \qquad   n=1,2, \ldots   \label{na_hier}
\ee
commute and any solution $\phi \in \A'$ solves the pKP hierarchy in $\A'$. 
}
\hfill $\square$
\vskip.2cm

Examples of WNA algebras are obtained as follows \cite{DMH06nahier}. 
Let $(\mathcal{A},\cdot)$ be any associative algebra and
$L, R \, : \, \mathcal{A} \rightarrow \mathcal{A}$ linear maps such that 
\be  
    [L,R] =0 \, , \qquad 
    L(a \cdot b) = L(a) \cdot b \, , \qquad 
    R(a \cdot b) = a \cdot R(b) \; .  \label{LR-rels}
\ee
It is convenient to write $L a$ and $a R$ instead of $L(a)$ and $R(a)$. 
Augmenting $\mathcal{A}$ with a constant element $\nu$ and setting 
\be
    \nu \circ \nu = 0 \, , \qquad \nu \circ a = L a \, , \qquad
    a \circ \nu = - a R \, , \qquad a \circ b = a \cdot b \, ,
\ee
leads to a WNA algebra $(\A,\circ)$ with $\A' = \mathcal{A}$, provided 
that there exist $a,b \in \mathcal{A}$ such that 
$aR \circ b \neq a \circ Lb$. The latter condition 
ensures that the augmented algebra is \emph{not} associative. 
As a consequence we obtain
\be
    \nu \ci_n a = L^n a \, , \qquad a \ci_n \nu = - a R^n \, ,
\ee
and
\be
    a \ci_n b = \sum_{k=0}^{n-1} a \, R^k \cdot L^{n-k-1} b \; .
\ee
Now (\ref{na_hier}) reads
\be
   \phi_{t_n} = L^n \phi - \phi R^n - \sum_{k=0}^{n-1} \phi \, R^k \cdot L^{n-k-1} \phi 
   \qquad n=1,2, \ldots   \; .        \label{na_hier2}
\ee
Introducing
\be
   \mathcal{W}(i,j) = L^i \phi R^j  \qquad \quad i,j=0,1,\ldots \, ,  \label{W_LR}
\ee
and acting on (\ref{na_hier2}) by $L^i$ from the left and by $R^j$ from 
the right, results in
\be
    \mathcal{W}(i,j)_{t_n} = \mathcal{W}(i+n,j) - \mathcal{W}(i,j+n) 
  - \sum_{k=0}^{n-1} \mathcal{W}(i,k) \cdot \mathcal{W}(n-k-1,j) \, , 
            \label{Weqs}
\ee
assuming that the partial derivatives $\pa_{t_n}$ commute with $L$ and $R$. 
\vskip.1cm

Suppose now that we have a solution $\{ \mathcal{W}(i,j) \}$ of the system (\ref{Weqs}). 
Then we can choose $\mathcal{A}$ as the associative algebra generated 
by this set (with product $\cdot$) and define the maps $L$ and $R$ via 
(\ref{W_LR}) and (\ref{LR-rels}). It follows (by use of the theorem) that 
$\phi = \mathcal{W}(0,0)$ solves the pKP hierarchy in $(\mathcal{A},\cdot)$. 
\vskip.1cm

In fact, in the first section we have shown how a subclass of solutions 
to the pKP hierarchy determines solutions $\{ \mathcal{W}(i,j) \}$ of (\ref{Weqs}) 
in the case where the product in $\mathcal{A}$ is given by $A \cdot B = AQB$ 
with a constant element $Q \in \mathcal{A}$. 
\vskip.1cm

In particular, we have seen that the system (\ref{Weqs_R}) is a special case 
of the hierarchy (\ref{na_hier}) of ordinary differential equations in a 
WNA algebra $\A$, which according to the above theorem determines solutions 
of the pKP hierarchy in $\A'$. 
An example of (\ref{Weqs_R}) appeared in \cite{Gils+Nimm07} and 
this will be the subject of the next section. 
\vskip.2cm

\noindent
\textbf{3. Solutions of noncommutative pKP hierarchies in terms of Wronski matrices.} \\
Let us now choose $\mathcal{A}$ as the algebra of $N \times N$ matrices with 
entries in a unital associative algebra $\mathcal{R}$ and product $A \cdot B = AQB$ 
with a constant $N \times N$ matrix $Q$. 
Let $\mathbf{e}_k$ be the $N$-component vector with all entries zero except for 
the identity element in the $k$th row. Choosing the rank 1 matrix 
\be
    Q = \mathbf{e}_N \mathbf{e}_N^T      \label{Q_rank1}
\ee 
(where ${}^T$ means taking the transpose), any solution $\phi$ of the 
pKP hierarchy in $(\mathcal{A},\cdot)$ determines via 
\be
     \varphi = \mathbf{e}_N^T \, \phi \, \mathbf{e}_N   \label{varphi}
\ee
a solution of the pKP hierarchy in $\mathcal{R}$. 
Choosing moreover
\be
     R = \Lambda  
       = \sum_{k=1}^{N-1} \mathbf{e}_k \mathbf{e}_{k+1}^T \, ,  \label{shift}
\ee 
which is the `left shift' ($\Lambda \mathbf{e}_1 =0$ and $\Lambda \mathbf{e}_k = \mathbf{e}_{k-1}$ for $k=2,\ldots, N$), the condition (\ref{CH_cond}) becomes
\be
    (I - e_N e_N^T) \pa(X) = \Lambda \, X  \; .
\ee
This tells us that $X$ is a Wronski matrix, i.e.
\be
    X = W(\Theta) 
  = \left(\begin{array}{cccc} \theta_1 & \theta_2& \cdots & \theta_N \\
    \pa(\theta_1) &\pa(\theta_2) & \cdots & \pa(\theta_N) \\
    \vdots & \vdots & \ddots& \vdots \\
    \pa^{N-1}(\theta_1) & \pa^{N-1}(\theta_2) & \cdots & \pa^{N-1}(\theta_N)
          \end{array}\right) \; .
\ee
with a row vector $\Theta = (\theta_1,\ldots,\theta_N)$ of 
elements of $\mathcal{R}$. 
We simply write $W$ instead of $W(\Theta)$ in the following. 
The next result is an immediate consequence of proposition~1. 
\vskip.2cm

\textbf{Proposition~3.} 
\emph{If $\theta_1, \ldots, \theta_N$ solve the linear heat hierarchy, i.e. 
$\Theta_{t_n} = \pa^n(\Theta)$, $n=1,2,\ldots$, and if the Wronski matrix $W$ 
is invertible, then
\be
     \phi = \pa(W) W^{-1}   \label{phi_Wronski}
\ee
solves the pKP hierarchy in the algebra of $N \times N$ matrices with 
entries in $\mathcal{R}$ and product $A \cdot B = AQB$ with $Q$ defined 
in (\ref{Q_rank1}). Furthermore, $\varphi$ defined in (\ref{varphi}) then 
solves the pKP hierarchy in $\mathcal{R}$. 
}
\vskip.2cm

According to section~1, $\phi$ given by (\ref{phi_Wronski}) determines a 
solution of the system (\ref{Weqs_R}). As a consequence, 
\be
  \mathcal{Q}(i,j) = - \mathbf{e}_N^T \, \mathcal{W}(i,j) \, \mathbf{e}_N 
    \qquad i,j=0,1,\ldots
\ee
solve the system
\be
   \mathcal{Q}(i,j)_{t_n} =  \mathcal{Q}(i+n,j) 
  - \mathcal{Q}(i,j+n) 
  + \sum_{k=0}^{n-1} \mathcal{Q}(i,k) \, \mathcal{Q}(n-k-1,j) \, ,
      \label{Qeqs}
\ee
which appeared in \cite{Gils+Nimm07}. 
 From the above definition we find immediately the following expression in terms 
of quasideterminants, 
\be
    \mathcal{Q}(i,j) 
  = - \mathbf{e}_N^T \, \pa^{i+1}(W) W^{-1} \mathbf{e}_{N-j} 
  = \left|\begin{array}{cc} W & \mathbf{e}_{N-j} \\
    \mathbf{e}^T_N \pa^{i+1}(W) & \fbox{0} \end{array} \right| \; .
\ee
(see \cite{Krob+Lecl95,GKLLRT95,GGRW05} for the notation). 
The authors of \cite{Gils+Nimm07} used Darboux transformations and properties 
of quasideterminants to derive these results. Knowing that (\ref{Qeqs}) 
holds, we can also refer directly to the arguments of section~2 
(instead of referring to the matrix solution $\phi$) to conclude that 
\be
      \varphi = - \mathcal{Q}(0,0)
  = - \mathbf{e}_N^T \, \pa(W) W^{-1} \mathbf{e}_N 
  = \left|\begin{array}{cc} W & \mathbf{e}_N \\
    \mathbf{e}^T_N \pa(W) & \fbox{0} \end{array} \right| 
\ee
solves the pKP hierarchy in $\mathcal{R}$.\footnote{This also solves the 
technical problems met by the authors of \cite{Gils+Nimm07} in their 
`direct approach'. For the use of computer algebra to perform 
computations of the kind considered in section 5 of \cite{Gils+Nimm07}, 
see also the appendix of \cite{DMH07Ricc}.}
\vskip.2cm

\textbf{Example.}
For $N=2$ we have
\be
    W = \left(\begin{array}{cc} \theta_1 & \theta_2 \\
        \theta'_1 & \theta'_2 \end{array}\right) \, , \qquad
    W^{-1} = \left(\begin{array}{cc} 
            (\theta_1 - \theta_2 \theta'_2{}^{-1}\theta'_1)^{-1} &
            (\theta'_1 - \theta'_2 \theta_2{}^{-1}\theta_1)^{-1} \\
            (\theta_2 - \theta_1 \theta'_1{}^{-1}\theta'_2)^{-1} &
            (\theta'_2 - \theta'_1 \theta_1{}^{-1}\theta_2)^{-1}\end{array}\right) \, ,
\ee
where $\theta'_k = \pa(\theta_k)$, and we need to assume that the inverses exist. 
This leads to
\be
   \phi = \left(\begin{array}{cc} 0 & 1 \\
        ( \theta''_1 - \theta''_2 (\theta'_2)^{-1} \theta'_1 )
        ( \theta_1 - \theta_2 (\theta'_2)^{-1} \theta'_1 )^{-1}
                      & \varphi
        \end{array}\right)  \, ,
\ee
where $1$ stands for the identity element of $\mathcal{R}$ and
\be
    \varphi = ( \theta''_1 - \theta''_2 \theta_2^{-1} \theta_1 )
              ( \theta'_1 - \theta'_2 \theta_2^{-1} \theta_1 )^{-1} \; .
\ee
Particular solutions of the heat hierarchy are\footnote{The exponentials are 
at least well-defined as formal power series in $\mathbf{t}$. Other solutions 
of the heat hierarchy are given by linear combinations of Schur polynomials 
in $\mathbf{t}$ with constant coefficients in $\mathcal{R}$. }
\be
    \theta_k = \sum_{j=1}^M A_{k,j} \, e^{\xi(\alpha_j)} B_{k,j} \qquad k=1,\ldots,N \, ,
\ee
with some $M \in \mathbb{N}$, constant elements $A_{k,j}, B_{k,j}, \alpha_j \in \mathcal{R}$, 
and $\xi(\alpha) = \sum_{m \geq 1} t_m \, \alpha^m$. 
In the `commutative case', one recovers $N$-soliton solutions in this way \cite{Hiro04}. 
\hfill $\square$
\vskip.2cm

\noindent
\textbf{4. Linearization of the system (\ref{Qeqs}).} \\
Let us introduce the infinite matrix 
$\mathcal{Q} = (\mathcal{Q}(i,j))$, and the corresponding 
shift operator 
\be
  \Lambda = \left(\begin{array}{cccc} 0 & 1 &   & \cdots     \\
                                      0 & 0 & 1 & \cdots     \\
                                    \vdots & \ddots & \ddots & \ddots   
            \end{array}\right) \; .  
\ee
Then (\ref{Qeqs}) can be expressed as a system of matrix Riccati equations, 
\be
   \mathcal{Q}_{t_n} = \Lambda^n \mathcal{Q} 
        - \mathcal{Q} (\Lambda^T)^n + \mathcal{Q} P_n \mathcal{Q}
   \qquad \quad  n=1,2,\ldots \, , 
\ee
with $P_1 = \mathbf{e}_1 \mathbf{e}_1^T$,  where now $\mathbf{e}_1^T = (1,0,\ldots)$, and
\be
    P_n = \sum_{k=0}^{n-1} (\Lambda^T)^k P_1 \Lambda^{n-k-1} \qquad \quad
    n=2,3, \ldots \; .
\ee
Such matrix Riccati equations are well-known to be linearizable. 
The corresponding linear system is 
\be
    \left(\begin{array}{c} \mathcal{X} \\ \mathcal{Y} \end{array}\right)_{t_n} =
    \left(\begin{array}{cc} \Lambda^T & P_1 \\ 0 & \Lambda \end{array} \right)^n
    \left(\begin{array}{c} \mathcal{X} \\ \mathcal{Y} \end{array} \right) 
    \qquad \quad  n=1,2,\ldots \, ,
\ee
which determines a solution of the matrix Riccati system via  
$\mathcal{Q} = \mathcal{Y} \mathcal{X}^{-1}$. Let us introduce
\be
     \hat{\mathcal{X}}_m  = \left\{ \begin{array}{l@{\qquad}l}
           \mathcal{X}_{-m}  & m<0 \\
           \mathcal{Y}_{m+1} & m \geq 0
          \end{array} \right.  \, ,
\ee
where $\mathcal{X}_m, \mathcal{Y}_m$, $m=1,2,\ldots$, are the rows of the 
matrices $\mathcal{X}, \mathcal{Y}$ (with entries in $\mathcal{R}$). 
In terms of the vector $\hat{\mathcal{X}} = (\hat{\mathcal{X}}_m)_{m \in \mathbb{Z}}$, 
this linear system takes the simple form
\be
     \hat{\mathcal{X}} _{t_n} = \hat{\Lambda}^n \hat{\mathcal{X}} 
       \qquad n=1,2, \ldots \, , 
\ee
with the two-sided infinite shift matrix $\hat{\Lambda}$. 
The solutions are given by $\hat{\mathcal{X}}(\mathbf{t}) = e^{\xi(\hat{\Lambda})}
\hat{\mathcal{X}}(0)$ with $\xi(\hat{\Lambda}) = \sum_{n \geq 1} t_n \, \hat{\Lambda}^n$. 
All this makes contact with Sato's formulation of the KP hierarchy as flows 
on an infinite-dimensional Grassmann manifold (see in particular 
\cite{Sato+Sato82,Sato89,Taka89}), 
but here the components of $\hat{\mathcal{X}}$ are taken from the (typically 
noncommutative) associative algebra $\mathcal{R}$. 
\vskip.2cm

\noindent
\textbf{5. Further remarks.} \\
In the transition from (\ref{na_hier2}) to (\ref{Weqs}) the linear maps 
$L$ and $R$ get hidden away and the infinite-dimensional shift operator 
enters the stage. This is made explicit in section~4, under the 
restrictions imposed in section~3. 
A system of the form (\ref{Weqs}), respectively (\ref{Qeqs}), already 
appeared in \cite{DNS89} (page 186), and in \cite{Taka89} (see in particular page 29) 
as a description of the (ordinary) KP hierarchy as Sato flows on the infinite 
(universal) Grassmann manifold (by vector fields associated 
with powers $\hat{\Lambda}^n$, $n=1,2, \ldots$, of the shift operator), 
see also \cite{MNT91}. 
Later it reappeared in \cite{FMP98,FMPZ00} as the ``Sato system" 
of the (ordinary) KP hierarchy.\footnote{See equation (7.1) in \cite{FMP98} 
and (2.8) in \cite{FMPZ00}. The correspondence is given by 
$\mathcal{W}(i,j) \mapsto - W^i_{j+1}$, respectively 
$\mathcal{Q}(i,j) \mapsto W^i_{j+1}$. 
The linearization presented in section~4 also appeared in these papers (where 
$\mathcal{R}$ is the commutative algebra of functions of $\mathbf{t}$). 
We believe that our presentation is somewhat improved. }
From a practical point of view, in particular when addressing exact solutions, 
it is in our opinion not of much help and it is more convenient and simpler 
to deal with (\ref{na_hier2}) 
(or the more general system (\ref{na_hier}), see also \cite{DMH06nahier,DMH07Ricc}). 
From a theoretical point of view, we have seen that the system (\ref{Weqs}) 
indeed has its merits. In particular it helped us bridging different approaches 
to solving a (noncommutative) KP hierarchy. 
\vskip.1cm

Our work makes evident that the choice of the shift operator 
($R=\Lambda$ in section~3) is rather special and there are others 
(see also \cite{DMH07Ricc}). 
We refer to the interesting discussion in \cite{Gekh+Kasm06} concerning 
the role of the shift operator in Sato theory and corresponding alternatives. 
\vskip.2cm

\noindent
\textbf{Appendix: A note on the (noncommutative) discrete pKP hierarchy.} \\
The (noncommutative) potential discrete KP (pDKP) hierarchy in an 
associative algebra $(\mathcal{A},\cdot)$ (see \cite{DMH06func,DMH07Ricc} and 
the references cited therein) is given by
\be
   \hat{\Omega}(\mu)^+ - \hat{\Omega}(\mu)_{-[\la]} 
 = \hat{\Omega}(\la)^+ - \hat{\Omega}(\la)_{-[\mu]} \, ,      \label{pDKP}
\ee
where 
\be
  \hat{\Omega}(\la) = \la^{-1} (\phi - \phi_{-[\la]}) 
      - (\phi^+ - \phi_{-[\la]}) \cdot \phi \, ,
\ee 
and $\phi = (\phi_k)_{k \in \mathbb{Z}}$, $\phi_k^+ = \phi_{k+1}$, with 
$\phi_k \in \mathcal{A}$. (\ref{pDKP}) is equivalent to
\be
    \hat{\Omega}(\la) = \vartheta^+ - \vartheta_{-[\la]}  \label{pDKP2}
\ee 
with some $\vartheta = (\vartheta_k)_{k \in \mathbb{Z}}$, $\vartheta_k \in \mathcal{A}$. 
Taking the limit $\la \to 0$ in (\ref{pDKP2}), results in 
\be
    \phi_{t_1} = (\phi^+ - \phi) \cdot \phi + \vartheta^+ - \vartheta \, , 
         \label{pDKP_BT}
\ee
by use of which (\ref{pDKP2}) is turned into the pKP system (\ref{pKP2}). 
Under the assumptions of proposition~1, we have $\vartheta = \phi R$ (see the 
remark in section~1), so that (\ref{pDKP_BT}) takes the form
\be
    \phi_{t_1} - (\phi^+ -\phi)(Q \phi +R ) = 0 \; .  \label{pDKP_BT2}
\ee
Assuming furthermore the Cole-Hopf restriction (\ref{CH}), we have
\be
    \phi_{t_1} = \pa^2(X) X^{-1} - \phi^2 
               = \pa^2(X) X^{-1} - \phi Q \phi - \phi(I-Q) \phi \; .
\ee
With the help of (\ref{CH_cond}), which is $(I-Q) \phi = R$, this becomes
$\phi_{t_1} = \pa^2(X) X^{-1} - \phi (Q \phi +R)$. 
Inserting this expression in (\ref{pDKP_BT}), leads to
$\pa^2 (X) X^{-1} =  \phi^+ (Q \phi +R) = \phi^+ \pa(X) X^{-1}$, which is
\be
    \phi^+ = \pa^2(X)(\pa(X))^{-1} \; .
\ee
Hence any invertible solution $X$ of the linear heat hierarchy, subject to 
(\ref{CH_cond}) (with any constant $Q,R$), determines a solution 
\be
    \phi_k = \pa^{k+1}(X) \, (\pa^k(X))^{-1}
\ee
of the pDKP hierarchy, restricted to non-negative integers $k$, provided that 
the inverse of $\pa^k(X)$ exists for all $k$. If the inverse $\pa^{-1}$ 
of $\pa$ and its powers can be defined on $X$, this extends to the whole lattice. 
In particular, `Wronski solutions' of pKP hierarchies as considered in section~3 
extend in this way to solutions of the corresponding pDKP hierarchies.

\end{document}